\documentclass[conference]{IEEEtran}

\usepackage{graphicx}
\usepackage{booktabs}
\usepackage{float}
\usepackage{afterpage}
\graphicspath{{pictures/}}
\usepackage{float}
\usepackage{titlesec}
\titlespacing*{\section}{0pt}{1.5ex}{0.5ex}  
\titlespacing*{\subsection}{0pt}{1ex}{0.5ex}
\title{\textbf{A Novel GPT-Based Framework for Anomaly Detection in System Logs}
}
\author{

    \IEEEauthorblockN{*Wenjie Yin}
    \IEEEauthorblockA{\\Hainan University \\ Haikou, China\\
    \textbf{2813350086@qq.com}
    }
    \and
    
    \IEEEauthorblockN{*Zeng Zhang}
    \footnote{These authors contributed equally to this work.}
    \IEEEauthorblockA{\\Hainan University \\ Haikou, China\\ \textbf{zz1up@hainanu.edu.cn}}
    \and
    \IEEEauthorblockN{Xiaoqi Li}
    \IEEEauthorblockA{\\Hainan University\\Haikou, China\\\textbf{csxqli@ieee.org}}
    
}
\date{\today}

\renewenvironment{abstract}
 {
  \begin{center}
  \bfseries \abstractname\vspace{-.5em}\vspace{0pt}
  \end{center}
  \list{}{\rightmargin\leftmargin}
  \item\relax}
 {\endlist}

 \usepackage{titlesec}

\titleformat{\section}{\normalfont\Large\bfseries}{\thesection}{1em}{}
\titleformat{\subsection}{\normalfont\large\bfseries}{\thesubsection}{1em}{}

\renewcommand{\thesection}{\arabic{section}}
\renewcommand{\thesubsection}{\arabic{section}.\arabic{subsection}}

\titleformat{\section}[block]
  {\normalfont\large\bfseries}{\thesection}{1em}{\MakeUppercase}

\usepackage{amsmath}
\usepackage[table]{xcolor}
\usepackage{multirow}
\usepackage{float} 
\usepackage{placeins}
\FloatBarrier 
\begin{document}

\maketitle
\footnotetext[1]{Zeng Zhang and Wenjie Yin contributed equally to this work.}
\begin{abstract}

Identification of anomalous events within system logs constitutes a pivotal element within the framework of cybersecurity defense strategies. However, this process faces numerous challenges, including the management of substantial data volumes, the distribution of anomalies, and the precision of conventional methods. To address this issue, the present paper puts forward a proposal for an intelligent detection method for system logs based on Generative Pre-trained Transformers (GPT). The efficacy of this approach is attributable to a combination of structured input design and a Focal Loss optimization strategy, which collectively result in a substantial enhancement of the performance of log anomaly detection. The initial approach involves the conversion of raw logs into event ID sequences through the use of the Drain parser. Subsequently, the Focal Loss loss function is employed to address the issue of class imbalance. The experimental results demonstrate that the optimized GPT-2 model significantly outperforms the unoptimized model in a range of key metrics, including precision, recall, and F1 score. In specific tasks, comparable or superior performance has been demonstrated to that of the GPT-3.5 API.

\end{abstract}

\section{Introduction}

The advent of information technology has precipitated a marked increase in the complexity of computer systems and network infrastructure. The substantial volumes of log data generated during system operation have become vital in network operations, security monitoring, and fault diagnosis\cite{zhu2019tools}. System logs are utilised extensively across servers, operating systems, network devices, databases, and security platforms as a system for the recording of events and state changes over time. The distinguishing characteristics of these logs are their high real-time nature and substantial information density. The efficient and accurate identification of potential anomalies within vast log data has become a core challenge in modern information security and intelligent operations management.

Conventional log analysis techniques predominantly rely on manually delineated rules and static thresholds, incorporating regular expression pattern matching, keyword filtering, and rudimentary statistical modelling\cite{fu2009execution}. While these approaches were once valuable, they have become increasingly inadequate when confronted with modern log data characterised by massive scale, complex structures, and ambiguous semantics. These include insufficient accuracy, poor generalisation capabilities, and high maintenance costs. Furthermore, anomalous events within logs typically exhibit scarcity and heterogeneity, manifesting in diverse forms and uneven distributions. Consequently, traditional analysis methods are rendered ineffective in detecting unseen anomalies and rare attack behaviors\cite{brown2020language}\cite{du2017deeplog}.

In recent years, significant progress has been made in the field of artificial intelligence, resulting in substantial advancements in deep learning and natural language processing techniques. The advent of the Transformer architecture\cite{vaswani2017attention} has precipitated revolutionary breakthroughs in the domain of natural language processing (NLP) through the utilisation of large-scale pre-trained language models. The GPT series models, with their formidable text modelling capabilities, have become indispensable tools for tasks such as anomaly detection, attack identification, and threat intelligence extraction\cite{radford2018improving}\cite{radford2019language}. It is noteworthy that OpenAI's GPT models have facilitated the identification of novel research prospects in the domain of log anomaly detection, a feat attributable to their remarkable aptitude in comprehending and generating language.

However, the direct application of these models to security scenarios still presents numerous challenges, such as class imbalance due to the scarcity of anomalous data, redundancy and inconsistent formatting in raw log texts, alongside engineering constraints including deployment costs and invocation efficiency\cite{du2017deeplog}\cite{zhang2019robust}. This research is situated within the broader context of the present study, to investigate the application of OpenAI's GPT series models in the identification of anomalies within structured system logs. The innovations concentrate on model input design, optimisation mechanisms, and evaluation methodologies to enhance the accuracy of log anomaly detection, bolster its practical utility, and advance the real-world application of pre-trained language models within the information security domain.

The contributions made by the present authors can be summarized as follows: firstly, a structured input method was proposed, based on event ID sequences, which significantly enhanced the model's comprehension of log semantics; secondly, the Focal Loss function was introduced, to effectively address class imbalance and improve anomaly detection performance; thirdly, an optimized GPT-2 fine-tuning framework was constructed, which achieved outstanding detection results under local deployment conditions; and fourthly, the method's efficacy and practical value were validated through comprehensive experimental evaluation\cite{he2016evaluation}.

\section{Background}

\subsection{GPT}

OpenAI's GPT models are constructed upon the Transformer architecture. The architectural design was proposed by Vaswani et al. in 2017\cite{vaswani2017attention}. The fundamental principle of this approach is predicated on self-attention mechanisms, a departure from the temporal dependency structure characteristic of traditional RNNs. This approach has been shown to achieve significant improvements in parallel computing capabilities and modeling long-range dependencies\cite{lee2023lanobert}\cite{lin2017focal}.

The Transformer's foundational structure incorporates multi-head self-attention mechanisms and feedforward neural networks. GPT utilizes a Decoder architecture, with a pivotal component consisting of stacked layers of Masked Multi-Head Attention. This layer utilises a masking technique to prevent the model from observing subsequent words, thereby attaining autoregressive prediction capabilities\cite{radford2018improving}. The Transformer's fundamental architecture encompasses multi-head self-attention mechanisms alongside feedforward neural networks. The core computational approach of the self-attention mechanism is illustrated as follows:

\begin{equation}
\text{Attention}(Q,K,V) = \text{softmax}\left( \frac{QK^{T}}{\sqrt{d_{k}}} \right)V
\end{equation}

In this context, $d_k$ denotes the key dimension, thereby serving as a scaling factor. Meanwhile, Q, K, and V correspond, respectively, to the query, key, and value matrices. In the GPT model, a decoder architecture is employed, comprising stacked layers of Masked Multi-Head Attention, which is identified as the most critical component. The configuration in question has been demonstrated to serve the purpose of preventing the model from observing subsequent words. The consequence of this phenomenon is the attainment of an autoregressive prediction effect within the model\cite{mcintosh2023harnessing}.

The GPT series models can be regarded as autoregressive language modelling systems based on the Transformer Decoder. The fundamental principle underpinning this approach entails preliminary training on extensive quantities of unsupervised text, to cultivate contextual relationships between words. This facilitates expeditious adaptation to subsequent tasks through the process of fine-tuning\cite{howard2018universal}. As OpenAI (2018) has stated, GPT-1 was pre-trained on the BooksCorpus dataset. The model under consideration has a size of 117 million parameters. This constitutes a pioneering example of the pre-training + fine-tuning paradigm in NLP\cite{radford2018improving}.

GPT-2 has been demonstrated to enhance model capacity to a considerable extent, with its largest variant exhibiting 1.5 billion parameters. The model has been trained to maximize conditional probability and has demonstrated exceptional zero-shot learning capabilities in unsupervised tasks, such as summary generation and question-answering systems\cite{vaswani2017attention}\cite{xu2024large}\cite{radford2019language}.

\begin{equation}
P(x) = \prod_{t = 1}^{T}{P\left( x_{t}\mid x_{1},x_{2},\ldots,x_{t - 1} \right)}
\end{equation}

GPT-3 has been further scaled to 175 billion parameters, incorporating the concept of few-shot prompting, and has achieved significant improvements across multiple tasks\cite{brown2020language}.

\subsection{Log Anomaly Detection}

Conventional intrusion detection systems are predominantly reliant upon the comparison of patterns and statistical rules, thereby attaining a high degree of accuracy in the identification of known attacks\cite{chandola2009anomaly}. However, the inherent limitations of these systems become evident when confronted with zero-day attacks and novel variant threats, resulting in their inability to meet the real-time and intelligent demands of contemporary network environments. In the early stages of research, the extraction of statistical features was predominantly conducted manually. The features in question included traffic volume, connection duration, and protocol type. Classical algorithms such as support vector machines, k-nearest neighbours, and decision trees were employed for the purpose of classification\cite{liu2008isolation}. However, these approaches encounter the 'curse of dimensionality' in high-dimensional spaces, resulting in suboptimal generalisation capabilities. Consequently, researchers introduced neural network models (e.g., multilayer perceptrons and convolutional neural networks) to enhance feature extraction\cite{meng2019loganomaly}. In recent years, sequence models, including recurrent neural networks and long short-term memory networks, have also been employed to model the time-dependent characteristics of network behaviour\cite{zhang2019robust}.

Advancements in natural language processing (NLP) have given rise to a surge of interest in the development of intrusion detection systems based on pre-trained language models. The employment of bidirectional transformer models, such as BERT and RoBERTa, in tasks such as anomaly log detection and system event prediction has demonstrated their capacity for robust generalisation and stability\cite{guo2021logbert}\cite{le2022log}. The GPT series models demonstrate considerable potential in the domains of semantic classification and attack description generation, leveraging their advanced language generation and comprehension capabilities\cite{huang2020hitanomaly}.

\subsection{Loss Function Optimisation}

In the context of security scenarios, researchers have identified the issue of class imbalance and have proposed corresponding improvement schemes. The Focal Loss algorithm was originally conceived as a solution to the tendency of traditional cross-entropy loss functions to overemphasise 'easily classifiable samples' in scenarios with severe class imbalance (Li et al., 2022). The fundamental principle underpinning this approach involves the reduction of the weight assigned to samples with higher predicted probabilities. This results in a shift of focus towards the challenging minority class samples\cite{lin2017focal}. This algorithm has been demonstrated to be particularly effective in the context of anomaly detection, where the proportion of anomalous samples is minimal\cite{chen2021experience}.

The formula for focal loss is as follows:

\begin{equation}
\text{fl} = - \alpha_{t}\left( 1 - p_{t} \right)^{\gamma}\log\left( p_{t} \right)
\end{equation}

When a model prediction attains a high level of confidence, it has been demonstrated that Focal Loss has the capacity to significantly reduce the sample's weight in the total loss. This results in a focus on anomalous samples with low prediction probabilities, which are prone to error. Conversely, cross-entropy treats all samples equally, resulting in an ineffective prioritisation of anomalous categories. Consequently, in system log analysis, replacing cross-entropy with Focal Loss substantially improves the model's recall and F1 score for minority class samples\cite{lin2017focal}.

\section{Intelligent Detection Method}
\subsection{Overall Architecture Design}

The present study proposes a high-performance, locally deployable system log anomaly detection solution. This solution is intended to address the current shortcomings in accuracy, efficiency, and controllability of large language models within the domain of information security applications\cite{yang2021semi}. The system architecture is predicated on the pre-trained language model GPT-2, thereby ensuring comprehensive design from theory to engineering through the collaborative operation of multiple modules, including input pre-processing, model training optimisation, and cross-model comparative evaluation.

\begin{figure}[H]
    \centering
    \includegraphics[width=1\linewidth]{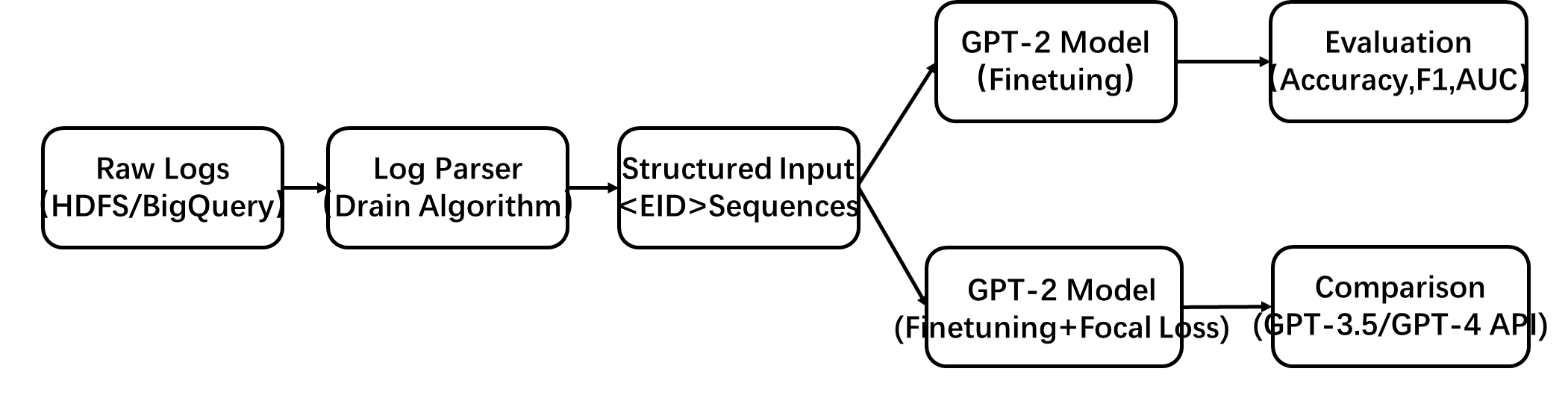}
    \caption{ Architectural Reference Diagram}
    \label{fig:placeholder}
\end{figure}

The system initially processes raw HDFS system logs as input. The aforementioned logs exhibit pronounced unstructured characteristics, incorporating fields such as timestamps, process IDs, log levels, component names, and log content\cite{xu2009detecting}. In order to enhance the model's comprehension of log semantics, researchers introduced a structured preprocessing strategy. The Drain algorithm\cite{he2017drain} was employed to map log entries to discrete event template IDs (Event IDs), which were then concatenated into fixed-length event sequences in chronological order.
\begin{table}[H]
\centering
\small
\caption{HDFS Log Dataset Characteristics}
\label{tab:dataset}
\begin{tabular}{@{}lll@{}}
\toprule
\textbf{Attribute} & \textbf{Value} & \textbf{Description} \\
\midrule
Data Source & LogHub & Standard HDFS dataset \\
Dataset Type & System Logs & Large-scale operations \\
Normal Logs & 97.07\% & Majority samples \\
Anomalous Logs & 2.93\% & Minority samples \\

Sample Size & 3,000 logs & Stratified sampling \\

Normal Samples & 2,912 logs & Original proportion \\

Anomalous Samples & 88 logs & Class imbalance \\
Sampling Method & Stratified & Representativeness \\
\bottomrule
\end{tabular}
\end{table}
During the process of data annotation, the label files provided by LogHub were utilised\cite{zhu2019tools}. Each event window was annotated as binary based on its classification within a known anomaly category, thereby constructing a high-quality training dataset. Training samples were configured to accept event sequences as input and anomaly detection as the target output, thereby transforming the problem into a binary classification task.

In terms of the model design, the pre-trained GPT-2 model provided by OpenAI was adopted, with a classification head added to transform the language modelling task into an anomaly detection task\cite{radford2019language}. In order to enhance the robustness of the model with regard to class-imbalanced data, the Focal Loss loss function was introduced with a view to guiding the model towards focusing more on learning samples with low confidence and those belonging to error-prone categories during the training process.

\subsection{Structured Log Input Design}

In the context of log anomaly detection tasks, system logs frequently manifest in unstructured natural language form, giving rise to challenges such as inconsistent formats, redundant information, and ambiguous semantics\cite{messaoudi2018search}. These factors have a detrimental effect on the ability of downstream models to model sequential patterns and anomalous behaviour, particularly in tasks that rely on pre-trained language models. The presence of redundant natural language inputs has been demonstrated to result in a significant number of invalid tokens, thereby impacting the efficiency of the model. Furthermore, these inputs have the potential to result in models neglecting to identify potential structural features of events.

In order to enhance semantic consistency during log sequence modelling and improve anomaly distinguishability, this study introduces a structured input mechanism based on event ID levels. The mechanism has been designed to transform raw log information into a series of controllable, discretised event sequences\cite{he2017drain}. Specifically, the Drain algorithm is employed for online parsing of raw logs to extract structural templates, assigning a unique event identifier (EventId) to each log template category. In the process of constructing training samples, a sliding window mechanism is employed to segment the structured log sequences into fixed-length windows, with each assigned a unique identifier (SessionId)\cite{du2017deeplog}. Concurrently, the anomaly\_label.csv file within the HDFS dataset is utilised to assign a binary classification label to each log window, indicating whether it contains anomalous events.

It is acknowledged that anomalous events constitute an extremely low proportion (approximately 2\% to 5\%) of actual system logs\cite{he2016evaluation}. In response to this, researchers introduced a stratified sampling strategy during the sampling phase. This ensured that the proportion of anomalous samples in the training set remained sufficiently high, thereby enhancing the model's robustness and generalisation capabilities during the learning phase. By controlling the sampling proportions across different categories, this strategy maintained category balance while ensuring the training data retained adequate representativeness.

The utilisation of structured input has been demonstrated to effect a substantial reduction in the number of input tokens, thereby enhancing processing efficiency and significantly improving the model's capacity to identify anomalous behaviour. A series of comparative experiments was conducted against a GPT-2 model devoid of structured input. These experiments demonstrated that utilising structured input resulted in a substantial enhancement of the F1-score and recall metrics, without compromising the model's accuracy. This lends further credence to the notion that the structured input approach is indeed valuable in the context of log anomaly detection scenarios.

\subsection{Model Optimisation Strategy}

In the context of log anomaly detection tasks, it is a common occurrence for the training data to demonstrate a significant imbalance across the various classes. In order to prevent models from diminishing their learning capacity for minority anomaly samples due to an overabundance of majority class instances during training, this study introduces the Focal Loss function to replace the standard cross-entropy loss function\cite{lin2017focal}. By adjusting the loss weighting mechanism, the model can allocate greater attention to difficult-to-classify anomaly samples, thereby enhancing overall detection performance.

In comparison with standard cross-entropy loss, Focal Loss significantly reduces the proportion of total loss attributed to samples when the model attains a reasonably high level of prediction accuracy. This approach has the effect of redirecting attention towards anomalous samples with low prediction probabilities and high error susceptibility. Conversely, cross-entropy loss treats all samples equally, failing to provide effective training focus for anomalous categories\cite{chen2021experience}.

In the model's specific implementation, secondary encapsulation of the Hugging Face Transformers framework was performed by inheriting the Trainer class, and the Compute Loss method was overridden to support loss calculation logic based on Focal Loss\cite{wolf2020transformers}. In order to enhance the stability of the training process, the AdamW optimiser was selected, in conjunction with learning rate warm-up and gradient clipping mechanisms. The purpose of these mechanisms is to effectively prevent gradient explosion and overfitting.

The incorporation of Focal Loss has been demonstrated to enhance the model's sensitivity and its capacity to accommodate minority class samples. Subsequent experimental analysis, through detailed ablation studies comparing models with and without Focal Loss, validated this strategy's significant effectiveness in improving metrics such as Recall and F1 score.

\subsection{Model Training and Inference Process}

The present study proposes a binary classification model for log sequences based on GPT-2, with fine-tuning on structured log sequences to enable the model to recognise system anomaly patterns\cite{radford2019language}. In order to ensure controllability, convergence, and robustness throughout the training process, systematic planning and design were undertaken for model initialisation, training parameter configuration, data partitioning methods, and inference procedures.

The training data comprises structured log samples, each containing a fixed-length sequence of event identifiers alongside corresponding anomaly labels. In order to prevent overfitting and to evaluate the model's generalisation capability, the training and validation sets were randomly partitioned at a 9:1 ratio. A stratified sampling strategy was employed to ensure consistent distribution of positive and negative samples across both sets, thereby enhancing the representativeness of the evaluation.

The pre-trained GPT-2 model was expanded through the incorporation of an additional classification head, which was utilised to transform the final token vector of the model's raw output into an anomaly prediction probability. The AdamW optimiser with weight decay was selected to accelerate convergence\cite{loshchilov2017decoupled}, with the learning rate set to $5 \times 10^{-5}$. Concurrently, a linear warm-up mechanism and gradient clipping were introduced to prevent gradient explosion. Furthermore, the Focal Loss function was employed as the loss function to ensure robust training stability even under conditions of extreme data imbalance.

The training process is divided into multiple epoch stages. Subsequent to the conclusion of each training iteration, the system promptly undertakes an evaluation of the key metrics on the validation set, and the model parameters are saved at the point at which these metrics are at their optimum. The approach is designed to achieve the training objective of minimising the loss function on the validation set.

During the training phase, the system automatically logs training loss, evaluation metrics, and learning curves for subsequent analysis of the impacts of different strategies. The results of such an analysis can then be used to generate confusion matrices and to compute metrics such as the area under the curve (AUC). The provision of inference outputs in a unified format alongside API model results is also a possibility. Upon completion of the training process, the system automatically saves the model weights to a predetermined directory, while simultaneously exporting the tokeniser configurations and training logs. This approach is instrumental in ensuring the reproducibility of experimental findings.

\section{Evaluation}
\subsection{Experimental Setup and Criteria}
The experiment utilised the standard HDFS log dataset provided by the LogHub platform\cite{zhu2019tools}. The dataset under consideration comprised both normal and anomalous logs, with the latter accounting for 2.93\% of the total. In view of the marked class imbalance in the log data, stratified random sampling was employed during the experiment. Samples were extracted from a total of 3,000 logs to ensure the ratio of abnormal to normal data in the sample matched that of the original dataset, thereby better approximating real-world application scenarios.

The experiments were conducted on a Windows 11 Professional operating system, utilising an Intel Ultra 7 processor, an NVIDIA RTX 4060 graphics card, and 32 GB of RAM to ensure stability throughout model training and evaluation. The development and training of the model were conducted within a Python 3.10 environment, utilising the PyTorch 2.0.1 and Hugging Face Transformers \cite{wolf2020transformers}.

During the training of the model, the learning rate was set at a uniform value of $2 \times 10^{-5}$, with a single epoch, an input batch size of 8, and the AdamW optimiser was utilised. Furthermore, the maximum sequence length was fixed at 128 tokens and the gradient accumulation step count at 4, thus serving as control variables to ensure consistent initial conditions across experimental groups.

The evaluation metrics encompass a multidimensional array of indicators, including, but not limited to accuracy, precision, recall, F1 score, and AUC. Accuracy is defined as the ratio of correctly predicted samples to the total sample size and is used to assess the overall correctness of the model in the identification. Precision is defined as the proportion of anomalous samples that are correctly identified as such, demonstrating the reliability of the model in predicting anomalies. It is imperative to recall, meanwhile, measures of the model's ability to detect all genuine anomalous samples, that is, the proportion of actual anomalies correctly identified.

Accuracy is defined as the ratio of the number of correctly predicted samples to the total number of samples. Its application is in the context of conducting an overall evaluation of the recognition accuracy of a model. :

Precision is defined as the proportion of genuinely anomalous samples among those classified as anomalous by the model. The reliability of the model in predicting anomalies is indicated by the function and its specific definition is as TABLE II.

\begin{table}[H]
\centering
\footnotesize
\caption{Evaluation Metrics}
\label{tab:metrics}
\begin{tabular}{@{}p{1.8cm}p{2cm}p{2.7cm}@{}}
\toprule
\textbf{Metric} & \textbf{Formula} & \textbf{Purpose} \\
\midrule
Accuracy & $\frac{TP + TN}{Total}$ & 
Overall correctness \\
\midrule
Precision & $\frac{TP}{TP + FP}$ & 
True anomaly rate \\
\midrule
Recall & $\frac{TP}{TP + FN}$ & 
Detection completeness \\
\midrule
F1-Score & $\frac{2 \cdot P \cdot R}{P + R}$ & 
Harmonic mean \\
\midrule
AUC & ROC Area & 
Class separation \\
\bottomrule
\end{tabular}
\end{table}

The definitions of TP, TN, FP, and FN referenced in the aforementioned formula are as follows: TP denotes the precise number of anomalous samples that the model correctly identifies as anomalous. The term FP is an abbreviation for False Positive, representing the number of normal samples erroneously classified as anomalous by the model. TN denotes the number of normal samples correctly identified as such by the model. The term FN, or False Negative, refers to the number of abnormal samples that are incorrectly classified as normal by the model.

\begin{table}[htbp]
\centering
\footnotesize
\caption{Model Training Parameters}
\label{tab:training_params}
\begin{tabular}{@{}p{2.5cm}p{1.5cm}p{2.5cm}@{}}
\toprule
\textbf{Parameter} & \textbf{Value} & \textbf{Rationale} \\
\midrule
Learning Rate & $2 \times 10^{-5}$ & Fine-tuning optimum \\
Epochs & 1 & Prevent overfitting \\
Batch Size & 8 & Memory efficient \\
Optimizer & AdamW & Weight decay \\

Max Length & 128 tokens & Context/efficiency \\

Grad. Accum. & 4 steps & Larger batch sim. \\
Loss Function & Focal Loss & Class imbalance \\
Architecture & GPT-2 + Head & Adaptation \\
Train/Val & 90\%/10\% & Standard practice \\
\bottomrule
\end{tabular}
\end{table}

\subsection{Experimental Results}

In order to validate the effectiveness of the proposed structured input and Focal Loss optimisation strategy on GPT-2 model performance, a detailed comparative analysis was conducted between the original GPT-2 model and the optimised GPT-2 model. The experimental results were interpreted across multiple dimensions, including model training curves, confusion matrices, and ROC curves. Utilising identical datasets and experimental conditions, alterations in the model's loss function and accuracy during training were documented. Cross-validation was employed during the training process, with performance evaluated on the validation set after each training iteration to ensure an objective assessment of accuracy.

A thorough analysis of the training process reveals that the optimised GPT-2 model converges more rapidly, exhibits a more pronounced reduction in loss function values, and ultimately achieves a lower convergence level. This finding emphasises the pivotal role of structured inputs and Focal Loss in enhancing the model's generalisation capabilities\cite{chen2021experience}.

\begin{figure}[H]
    \centering
    \includegraphics[width=1\linewidth]{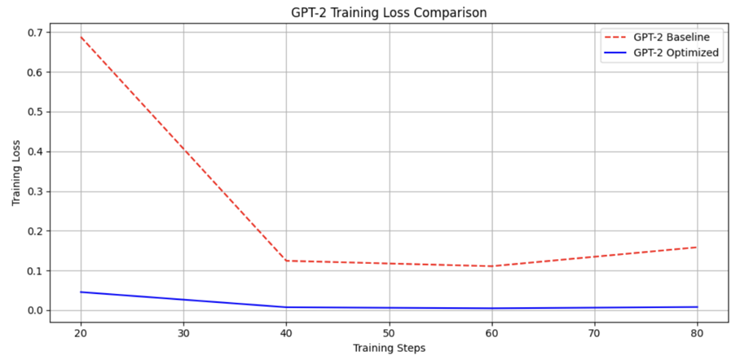}
    \caption{Training Loss Curves for the Original and Optimised GPT-2 Models}
    \label{fig:placeholder}
\end{figure}
A series of analyses has demonstrated that the optimised GPT-2 model exhibits considerable advantages in the domain of anomaly detection tasks. On the one hand, the model effectively detects and precisely localises anomalies with high recall rates; on the other hand, it ensures the accuracy of anomaly identification, achieving precision approaching 100\%. Substantial enhancements have been accomplished across a range of comprehensive performance metrics, encompassing overall accuracy, recall rates, and F1 scores.

\subsection{Experimental Design and Effect Verification for Melting}

In order to provide further validation of the practical efficacy of the two key optimisation strategies proposed in this study (structured input and focus loss), a series of ablation experiments was designed. By controlling variables, certain strategies were stripped or replaced in order to systematically analyse the independent contribution of each optimisation point towards enhancing model performance.

The experiments compared three model groups: Model A (Baseline) employed the unmodified GPT-2 model, utilising unstructured raw log text as input and training with a standard cross-entropy loss function; Model B used structured EID sequences as input but retained the cross-entropy loss function; Model C, building upon Model B, further incorporated Focal Loss during training, constituting the final optimised model version proposed in this study.

All experiments employed identical training data (3,000 entries selected via stratified sampling) while maintaining consistent model hyperparameters and training strategies to eliminate interference from other variables. The findings of the study demonstrate that the incorporation of structured input has a substantial impact on the model's capacity to identify anomalous samples. In the context of a loss function condition that was identical for both models, Model B achieved an F1 score of 0.574 by altering the input format. This result demonstrates that the utilisation of structured template inputs enables the model to capture anomalous features within logs more effectively. The subsequent introduction of Focal Loss (Model C) yielded a precision of 1.000 and an F1 score of 0.800, indicating a substantial improvement in identifying imbalanced samples.

The incorporation of structured inputs has been demonstrated to enhance the model's capacity to model log semantics. It is evident that focal loss serves to further accentuate the model's predilection for the minority anomalous log categories. The integration of these two approaches has been demonstrated to substantially enhance the model's overall performance in practical system log anomaly detection tasks\cite{lin2017focal}.

\begin{figure}[htbp]
    \centering
    \includegraphics[width=1\linewidth]{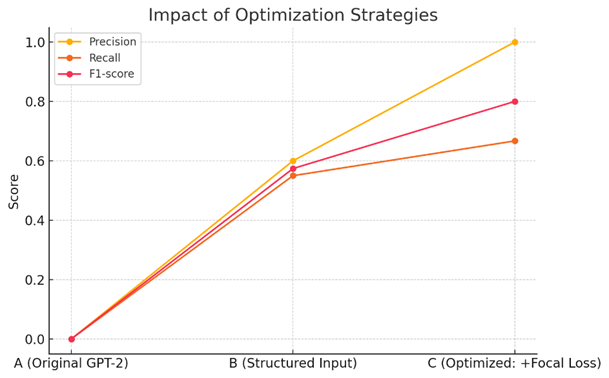}
    \caption{Comparative Analysis of Model Optimization Approaches}
    \label{fig:placeholder}
\end{figure}
\subsection{Comparative Analysis with GPT-3.5/GPT-4}

In order to provide further validation of the practical effectiveness of the optimised local GPT-2 model in system log anomaly detection tasks, a comparison was made against mainstream large language model APIs (GPT-3.5-turbo and GPT-4) under identical data conditions\cite{achiam2023gpt}. The performance of various models in anomaly detection was assessed by means of a multifaceted evaluation that encompassed the standardisation of test samples, input formats, and evaluation metrics. This methodological approach facilitated a comprehensive analysis of the models' respective strengths and limitations.

The experiment involved the hierarchical extraction of 300 samples from structured HDFS log data, encompassing approximately 9\% anomalous entries as the test set. In accordance with the standardisation of input formats, the samples were entered into the locally optimised GPT-2, GPT-3.5, and GPT-4 models, respectively. Each model was instructed to determine whether each log entry constituted an anomaly. In order to ensure a fair comparison, the API models operated under a zero-shot setting using a unified prompt.

The comparison revealed that both GPT-3.5 and GPT-4 demonstrated strong recall performance, achieving 0.90 and 1.00, respectively, indicating a degree of anomaly detection capability. However, the precision of these models was found to be suboptimal at 0.49 and 0.50, suggesting that they may generate a higher number of false positives, incorrectly flagging normal logs as anomalies. In contrast, the locally optimised GPT-2 model not only achieved 100\% precision but also demonstrated a significant improvement in recall, reaching 0.667. The final F1-score of 0.800 for the model in question was significantly higher than that of competing models, with the model demonstrating superior balance and practicality.

\begin{table}[htbp]
\centering
\footnotesize
\caption{Model Performance Comparison (Compact)}
\label{tab:model_performance_compact}
\small
\begin{tabular}{@{}lrrrr@{}}
\toprule
\textbf{Model} & \textbf{Acc.} & \textbf{Prec.} & \textbf{Rec.} & \textbf{F1} \\
\midrule
GPT-2 (Original) & 0.970 & 0.000 & 0.000 & 0.000 \\

GPT-2 (Optimized) & {0.990} & {1.000} & 0.667 & {0.800} \\
GPT-3.5-turbo & 0.487 & 0.493 & {0.900} & 0.637 \\
GPT-4 & 0.500 & 0.500 & {1.000} & 0.667 \\
\bottomrule
\end{tabular}
\end{table}

  From a deployment perspective, API models demonstrate robust contextual understanding and generalisation capabilities\cite{kaddour2023challenges}. However, during the implementation of specific tasks, certain instabilities and uncontrollable factors persist, primarily including:

The output displays a variety of stylistic variations. Despite explicit prompt requirements for a specific format, API responses may still contain explanatory content or redundant information, which complicates result parsing.

The reasoning process demonstrates inconsistencies. It has been demonstrated that the invocation of the API at differing times for samples that are identical may result in the occurrence of fluctuating results\cite{wang2023robustness}.

Furthermore, API models entail higher invocation costs and exhibit strong network dependency, requiring an internet connection for operation. This renders them vulnerable to network conditions and billing policies, thus rendering them unsuitable for scenarios involving sensitive data.

In contrast, the locally optimised GPT-2 model incorporates structured inputs and Focal Loss optimisation during training, fully leveraging the structural characteristics of log sequences\cite{he2017drain}. The model employs a balanced processing strategy for anomaly categories, thereby enhancing model stability and ensuring highly controllable prediction outcomes. This model is well-suited to offline environments, showcasing enhanced deployment flexibility in information security-related scenarios and possessing substantial engineering application value.

In summary, for specific system log anomaly detection tasks, the locally optimised GPT-2 model developed in this study delivers performance comparable to or surpassing mainstream large model APIs while maintaining high efficiency. The former demonstrates particular advantages in terms of accuracy and practicality. This finding serves to substantiate the practical innovation and application potential of this research within the domain of information security.

\section{Related Work}

\subsection{Traditional Log Anomaly Detection Methods}

Conventional intrusion detection systems are predominantly reliant on the comparison of signatures and statistical rules, exhibiting a high degree of accuracy in the identification of known attacks. However, when confronted with zero-day attacks and novel variant threats, these systems demonstrate evident limitations, resulting in suboptimal performance in meeting the real-time and intelligent demands of contemporary network environments (Smith, 2023)\cite{chandola2009anomaly}. Consequently, intelligent intrusion detection systems based on machine learning and deep learning have progressively become the mainstream research direction\cite{zhang2019robust}.

In the early stages of research, the focus was predominantly on manually extracted statistical features, including traffic volume, connection duration, and protocol type. The subsequent classification of these features was undertaken through the utilisation of classical algorithms, including support vector machines, k-nearest neighbours, and decision trees\cite{liu2008isolation}. However, these approaches encounter the 'curse of dimensionality' in high-dimensional spaces, resulting in suboptimal generalisation capabilities. Consequently, researchers began incorporating neural network models---such as multilayer perceptrons and convolutional neural networks---to enhance feature extraction capabilities\cite{meng2019loganomaly}.

In recent years, sequence models, including recurrent neural networks and long short-term memory networks, have been employed to model the temporal dependencies of network behaviour. Recent advancements in the domain of natural language processing (NLP) have given rise to research in intrusion detection systems that are based on pre-trained language models. The employment of bidirectional transformer models, such as BERT and RoBERTa, has been demonstrated in a variety of applications, including anomaly log detection and system event prediction (Smith, 2023)\cite{guo2021logbert}. These models have demonstrated noteworthy capabilities in generalisation and stability, thereby highlighting their efficacy in a range of tasks.

\subsection{Deep Learning-Based Log Analysis}

Transformer-based models have been demonstrated to attain satisfactory performance in the identification of log anomalies\cite{le2022log}. Nevertheless, the principal emphasis of their design is on NLP tasks, a factor which may impede their capacity to adapt to the compact structure and relatively sparse events that characterise system log corpora. In scenarios where the number of anomaly samples is low (i.e., below 5\%), standard loss functions such as cross-entropy are ineffective for training the model's positive class\cite{chen2021experience}.

In consideration of the structural and semantic idiosyncrasies that are an inherent feature of log data, the efficacy of pre-trained models is contingent upon the successful implementation of effective task transfer and structural adaptation, with the objective of optimising their performance advantages\cite{huang2020hitanomaly}. The present study proposes an optimisation scheme for pre-trained language models to better accommodate log semantic features. Firstly, the scheme involves the structural transformation of log inputs through the process of event template abstraction. This is with a view to reducing redundant information interference and enhancing input consistency. Secondly, the Focal Loss function is introduced, which is suitable for imbalanced classification, with the objective of improving model performance in the context of minority class anomaly detection tasks\cite{lin2017focal}.

\subsection{Applications of Large Language Models in the Security Domain}

In comparison with preceding research, this paper not only focuses on improving model accuracy but also emphasises the model's inference efficiency, API model comparison capabilities, and security and engineering feasibility in integrated application scenarios under actual deployment conditions\cite{yang2021semi}. At present, domestic research is investigating numerous dimensions concurrently. These include model structure optimisation, data privacy protection, and explainability. The objective of this exploration is to enhance the engineering application of large language models in the security domain.

The technical approach of this study aligns closely with prevailing research directions in security-oriented large language models, further validating the substantial application potential of such models in domains such as automated threat response. A comprehensive review of extant research has been undertaken, and it is concluded that this provides a critical theoretical and technical foundation for the present work. Nevertheless, there is still scope for improvement in areas including input representation, loss function design, and training mechanism optimisation.

\section{Discussion}

We focus on intelligent system log detection methods based on GPT, particularly system log anomaly detection. We examine application pathways and optimisation approaches for pre-trained language models in practical security scenarios in depth. Through a combination of theoretical analysis and experimental practice, we develop an innovative, adaptable, and valuable methodology for anomaly detection.

We implemented a structured strategy for modelling log inputs. The log parsing process converts raw system logs into a unified format using templates and event ID mappings\cite{he2017drain}. Transforming unstructured log data into event sequences allows us to analyse system activities systematically. This enhancement allows the model to capture context dependencies and semantic rules, paving the way for stable language model training and efficient inference.

We set out the theory of Focal Loss as a means of optimising detection performance. In addressing the prevalent class imbalance issue in anomaly detection, the traditional cross-entropy loss function was substituted with Focal Loss\cite{lin2017focal}. This approach prioritises focus on hard-to-classify samples, thereby improving the model's recall and F1 score for anomalous classes and enhancing its reliability in high-risk detection scenarios.

\section{Conclusion}
The construction and training of an optimised GPT-2 fine-tuning model is achieved through fine-tuning based on GPT-2. The employment of structured inputs and loss function optimisation resulted in a synergistic enhancement of the model's performance. A comparison of this iteration with the original model demonstrated superior performance across multiple evaluation metrics, thus validating the effectiveness of the proposed methodology. Structured training samples are constructed using authentic HDFS log data (zhu2019tools), thereby enabling a systematic comparison between the optimised model, the original model, and the API model. The performance contributions of each key module are assessed through a series of ablation experiments. The experimental results fully validate the feasibility and practical value of the research methodology.

We propose innovative and effective improvement strategies across three critical aspects: input representation, loss optimisation, and training mechanisms. The text establishes a theoretical foundation for implementing pre-trained language models in the domain of information security, while outlining concrete engineering implementation pathways. This research makes a valuable contribution to the field of security detection techniques. Moreover, it establishes a foundation for the prospective integration of large models and cybersecurity.

In summary, this paper conducts an in-depth investigation into intelligent anomaly detection based on language models. The research encompasses a number of aspects, including the construction of theoretical models, the design of algorithmic mechanisms, the implementation of systems, and the validation of experiments. This provides a comprehensive technical framework and experimental basis for this technology. The findings demonstrate that through rational input structure design and loss function optimisation, pre-trained language models can be effectively applied to system log anomaly detection tasks, thus offering robust technical support for cybersecurity protection.

\bibliographystyle{ieeetr}
\bibliography{sample}

\end{document}